\documentclass[aps,pre,twocolumn,showpacs,floatfix]{revtex4}
\usepackage{graphicx}
\usepackage{bm}
\usepackage{amsmath}
\usepackage[centerlast]{subfigure}
\newif\iffigs
\figstrue   
\iffigs
\fi
\def\drawing #1 #2 #3 {
\begin{center}
\setlength{\unitlength}{1mm}
\begin{picture}(#1,#2)(0,0)
\put(0,0){\framebox(#1,#2){#3}}
\end{picture}
\end{center}}

\begin{document}
\title{Thin front propagation in random shear flows} 

\author{M. Chinappi$^{1}$, M. Cencini$^{2,3}$ and A. Vulpiani$^{3,4}$
} \affiliation{ $^1$ Dipartimento di Meccanica e Aeronautica,
Universit\`a di Roma "la Sapienza", Via Eudossiana 18, I-00184 Roma,
Italy.\\ $^2$ ISC-CNR via dei Taurini, 19 I-00185 Roma, Italy\\ $^3$
SMC-INFM c/o Dipartimento di Fisica, Universit\`a di Roma ``La
Sapienza'', p.zzle Aldo Moro, 2 I-00185 Roma, Italy\\ $^4$Dipartimento
di Fisica and INFN, Universit\`a di Roma ``La Sapienza'', p.zzle Aldo
Moro, 2 I-00185 Roma, Italy} \date{\today}

\begin{abstract}
Front propagation in time dependent laminar flows is investigated in
the limit of very fast reaction and very thin fronts, i.e.  the
so-called geometrical optics limit.  In particular, we consider fronts
evolving in time correlated random shear flows, modeled in terms of
Ornstein-Uhlembeck processes. We show that the ratio between the time
correlation of the flow and an intrinsic time scale of the reaction
dynamics (the wrinkling time $t_w$) is crucial in determining both the
front propagation speed and the front spatial patterns.  The relevance
of time correlation in realistic flows is briefly discussed in the
light of the bending phenomenon, i.e. the decrease of propagation
speed observed at high flow intensities.
\end{abstract}

\maketitle

\begin{figure}[t!]
  \iffigs \includegraphics[scale=.6]{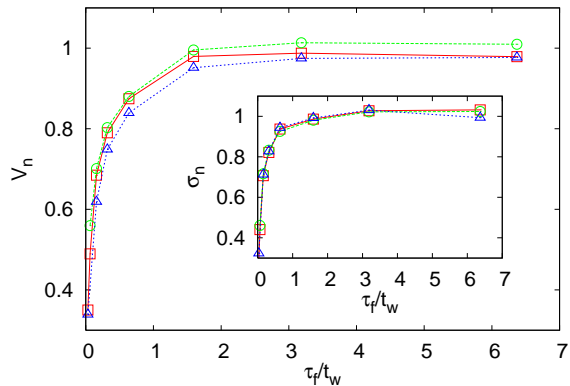}
  \else \drawing 70 40 {Fig4.eps} \fi
  \caption{\label{fig:4} Normalized velocity $V_n=(V_f-v_0)/\langle
\vert U \vert \rangle$ as a function of $\tau_f/t_w$ for $v_0=0.2$ and
$U_{rms}=2\sqrt{2}$ (circles, green online), $U_{rms}=\sqrt{2}$
(squares, red online) and $U_{rms}=1/\sqrt{2} $ (triangles, blue
online). The inset displays the normalized variance $\sigma_n$ in the
same cases.  The resolution used goes from $N_y=3000$ (for the lowest
value of $\tau_f$) up to 800 (for the highest one).}
\end{figure}

Front propagation in fluid flows is a problem relevant to many areas
of science and technology ranging from combustion technology
\cite{W85} to chemistry \cite{chem} and marine ecology \cite{bio}.  In
the last years several theoretical
\cite{ABP00,Const,KBM01,ACVV01,ACVV02,NX03,CTVV03} and experimental
\cite{LMRS03,PH04,SS05,LMRS05} works studied chemically reactive
substances stirred by laminar flows. This problem, though considerably
simpler than the case of turbulent flows \cite{W85}, is non trivial
and displays a very rich and interesting phenomenology.  We mention
here the front speed locking phenomenon in time dependent cellular
flows, which was numerically and theoretically found in
Ref.~\cite{CTVV03} and then experimentally observed in
Ref.~\cite{SS05}. Another interesting example is represented by the
theoretical studies on time periodic shear flows \cite{KBM01,NX03} and
the recent experimental work \cite{LMRS05} which study aqueous
reactions in periodically modulated Hele-Shaw flows.\\\indent Laminar
flows are interesting also because they constitute a theoretical
laboratory to study some problems which can be encountered in more
complex (turbulent) flows.  For instance, this is the case of time
correlations \cite{D99,A00} that are believed to be very important in
determining the bending of turbulent premixed flame velocity when the
intensity of turbulence is increased (see \cite{Ronney} for a
discussion about this problem).  Actually for bending several
mechanisms have been proposed like reaction quenching \cite{quench},
dynamics of pockets of material which did not react left behind
\cite{pockets} and finally time correlations \cite{D99,A00}.\\\indent
The aim of this paper is to investigate the role of time correlations
in the propagation of reactions in random shear flows. In particular,
we shall consider the problem in the so-called geometrical optics, or
Huygens regime \cite{Peter} that is realized in the case of very fast
reactions, taking place in very thin regions.  As in
Refs.\cite{D99,A00}, we neglect possible back-effects of the
transported reacting scalar on the velocity field, i.e. we treat the
problem in the context of passive reactive transport. The latter
assumption is justified for dilute aqueous auto-catalytic reactions
and more in general for chemical reaction with low heat release.  It
should be also remarked that in the chosen framework pockets cannot be
generated due to shear geometry and quenching of reaction cannot
happen due to the choice of working in the geometrical optics
limit. Therefore, the case under consideration allows us to focus on
the effects due to time correlations solely.\\\indent Let us start to
introduce our problem by shortly recalling the main equations. Since
we consider premixed reactive species, the simplest model consists in
studying the dynamics of a scalar field $\theta(\bm x,t)$ representing
the fractional concentration of the reaction's products ($\theta=1$
inert material, $\theta=0$ fresh one and $0<\theta<1$ coexistence of
fresh material and products).  The evolution of $\theta$ is ruled by
the advection-reaction-diffusion equation:
\begin{equation}
\partial_t \bm u+\bm u\cdot\bm\nabla \theta=D\Delta\theta+
\frac{f(\theta)}{\tau_r}\,,
\label{eq:ard}
\end{equation}
where $\bm u$ is a given velocity field (incompressible $\bm
\nabla\cdot\bm u= 0$ through this paper). The $f(\theta)$ (that is
typically a nonlinear function with one unstable $\theta=0$ and one
stable $\theta=1$ state) models the production process occurring on a
time-scale $\tau_r$.\\\indent Eq.~(\ref{eq:ard}) may be studied for
different geometries and boundary conditions. In this work we consider
an infinite two-dimensional stripe along the x-direction with a
reservoir of fresh material on the right, inert products on the left
and periodic boundary conditions in the transverse direction (which
has size $L$).  In particular, we shall be concerned with the
concentration initialized as a step, i.e. $\theta(x,y,0)=1$ for $x\leq
0$, and zero otherwise.
With this geometry a front of inert material (stable phase) propagates
from left to right with an instantaneous velocity which can be defined
as:
\begin{equation}
v_f(t) = \frac{1}{L} \int_\mathcal{D} d \bm x \;\partial_t \theta(\bm x,t)\,,
\label{eq:burningrate}
\end{equation}
more precisely this is the bulk burning rate~\cite{Const} (integration
is over the entire domain $\mathcal{D}$). Most of the theoretical
studies aim to predict the dependence of the average front speed
$V_f=\langle v_f \rangle$ on the details of the velocity field.  Very
important are of course also the propagation speed fluctuations; one
would like to predict how these are related to the fluid velocity
fluctuations. These are in general very difficult issues, but very
important in technological applications, where one has to project the
reactor geometry and flow characteristics.  Definite answers about the
reaction propagation exists only in particular conditions, e.g. when
the flow is motionless ($\bm u =0$) and under rather general
hypothesis on the production function $f(\theta)$  it is possible to
show that the reaction asymptotically propagates with a velocity $v_0$
within the bounds (see Ref.~\cite{saarloos} for an exhaustive review):
\begin{equation}
2\sqrt{\frac{Df'(0)}{\tau_r} } \leq  v_0 
\leq 2\sqrt{\frac{D}{\tau_r} \sup_{\theta} \left\{ \frac{f(\theta)}{\theta}
\right\}}\,,
\label{eq:aw}
\end{equation}
where $f'$ indicates the derivative, and the thickness of the reaction
zone varies as $\xi \propto \sqrt{D\tau_r}$.  For a wide class of
reaction terms $f$, such as the autocatalytic reaction dynamics,
$f(\theta)=\theta(1-\theta)$, and more in general for convex functions
($f^{''}(\theta)<0$) one can prove that $v_0=2\sqrt{Df'(0)/\tau_r}$
exactly.  In the presence of a velocity field $\bm u$, generally one
has that the speed $V_f$ is larger than the bare velocity
$v_0$. Specifically here we consider the limit in which the reaction
is much faster than the other time-scales of the problem, formally
this regime is reached when $\tau_r \to 0$ and $D\to 0$ but with
$D/\tau_r =const$ such that the bare propagation velocity $v_0=
2\sqrt{D/\tau_r}$ is finite and well defined, while the reaction zone
thickness shrinks to zero ($\xi \to 0$) \cite{Peter}, where for the
sake of notation simplicity we posed $f'(0)=1$.. It should be noted
that this regime, also called geometrical optics limit is commonly
encountered in many applications \cite{Peter}.  In this limit, being
sharp ($\xi\to 0$), the front dynamics can be described in terms of
the evolution of the surface (line in $2d$) which divides inert
($\theta=1$) and fresh ($\theta=0$) material.  The effect of the flow
is thus to wrinkle the front increasing (in two dimensions) its length
$\mathcal{L}_f$ and, as a consequence of the relation \cite{W85,Peter}
\begin{equation}
v_f = \frac{v_0 \mathcal{L}_f}{L}\,,
\label{eq:relation}
\end{equation}
(where $L$ is the length of a flat front in the absence of fluid
motion) its propagation velocity, i.e. $v_f>v_0$. Quantifying such an
enhancement is one of the main goals of, e.g., the community
interested in combustion propagation \cite{W85}. It should be also
remarked that the presence of complicated flow has also an important
role in the generation of patterns, i.e., front spatial
structures.\\\indent From a formal point of view the evolution of
$\theta$ can be recast in terms of the evolution of a scalar field
$G({\bf r},t)$, where the iso-line (in 2d) $G({\bf r},t)=0$ represents
the front, i.e. the boundary between inert ($G>0$) and fresh ($G<0$)
material.  $G$ evolves according to the so-called $G$-equation
\cite{Peter,EMS95}
\begin{equation}
\partial_t G +\bm u\cdot \bm \nabla G=v_0 |\bm \nabla G|\,.
\label{eq:Geq}
\end{equation}
The analytical treatment of this equation is not trivial, and even in
relatively simple cases (e.g., shear flows) numerical analysis is
needed. Also on the numerical side solving (\ref{eq:Geq}) is a non
trivial issue, indeed the presence of strong gradients usually requires
the regularization of (\ref{eq:Geq}) through the introduction of a
diffusive term (see e.g. \cite{Ald}). Here, following
Ref.~\cite{CTVV03}, we adopt a Lagrangian integration scheme the basic
idea of which is now briefly sketched.\\\indent
\begin{figure}[b!]
  \iffigs \includegraphics[scale=.45]{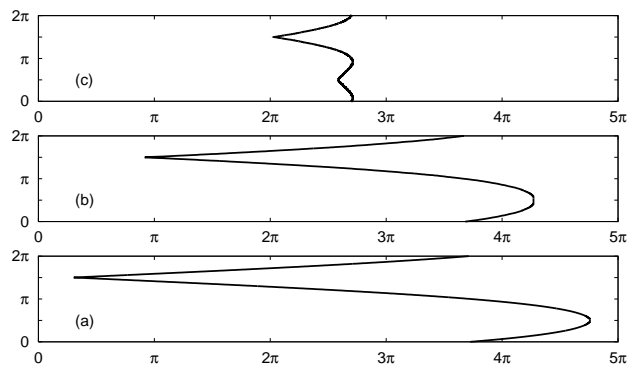}
  \else \drawing 70 90 {Fig1.eps} \fi
  \caption{\label{fig:1} Typical front patterns for a stationary shear
flow (a), time correlated shear flow with $\tau_f=200$ (b) and with
$\tau_f=2$ (c). In the stationary case $U=1/\sqrt{2}$, while in the time
correlated case we set $U_{rms}=\sqrt{2}$. In all cases the bare
velocity is $v_0=0.2$.  For (a) and (b) we used $N_y=800$ grid
points and $N_y=3000$ for (c). Here and in the following figures $L=2\pi$.}
\end{figure}
First of all let us introduce the type of flow we are interested in,
we consider shear flows that can be written as
\begin{equation}
\bm u=(U(t)g(y),0)\,,
\label{eq:shear}
\end{equation}
being $g(y)$ the functional shape of the flow (here $g(y)=\sin (2\pi
y/L)$) and $U(t)$ its intensity.  The domain of integration is chosen
as a stripe $[0:N\,L]\times[0:L]$, where $N$ (typically in the range
$5-20$) is the maximum number of cells of size $L$ in the
$x$-direction that are used in the integration (the number should be
fixed according to the front width).  The number of cells $N$ is
dynamically adjusted. In particular, after the propagation sets in a
statistically stationary regime, while the front propagates the cells
on the left that are completely inert with $\theta=1$ are eliminated
by the integration domain.  On the right side we retain only a finite
number (which depends on the maximum allowed speed) of cells with
fresh material $\theta=0$.  The domain is discretized and the value of
$\theta$ in each point of the lattice is updated with the following
rule. At each time step, each grid point $\bm r_{n,m}=(x_n,y_m)$ is
backward integrated in time according to the advection by the flow
$d\bm r/dt=-\bm u$. Once the point $\bm r'$ that will arrive in $\bm
r_{n,m}$ at time $t$ is known, $\theta(\bm r_{n,m},t)$ is set to $1$
if in a circle centered in $\bm r'$ and having radius $v_0dt$ there is
at least one grid point with $\theta=1$. This is a straightforward way
to implement the Huygens dynamics.
The algorithm works as soon as $v_0dt$ is sufficiently larger than the
spatial discretization $dx=dy=L/N_y$ (where $N_y$ is the number of
grid points in the $y$-direction, and $N_x=N\,N_y$). For a detailed
description of the algorithm see the Appendix in Ref.~\cite{CTVV03}.
\begin{figure}[t!]
\iffigs \includegraphics[scale=.55]{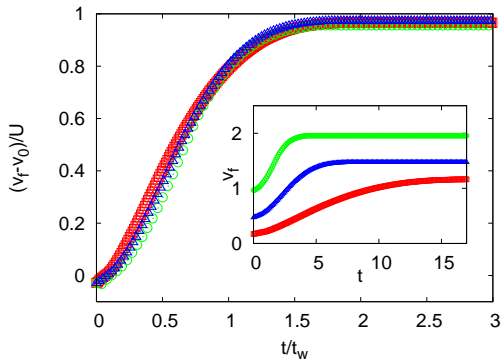}
\else \drawing 70 40 {Fig2.eps} \fi
\caption{\label{fig:2} Normalized velocity $V_n=(v_f(t)-v_0)/U$ vs
$t/t_w$ for: $v_0=1$, $U=1$ (circles, green online), $v_0=0.2$, $U=1$
(squares, red online) and $v_0=0.5$, $U=1$ (triangles, blue
online). The corresponding $t_w$'s were numerically computed as
$W_f^\ast/2U$ obtaining $2.6$, $9.48$ and $18.96$, respectively
($W_f^\ast$ is estimated by counting the number of pixel in the
border between inert and fresh material). The inset shows the unscaled
results.  The resolution used is $N_y=800$.}
\end{figure}
For a stationary shear flow, i.e. $U(t)=U$, by means of
simple geometrical reasonings one can show that at long times the
front evolves with velocity \cite{ABP00}:
\begin{equation} 
V_f=v_0+U\sup_{y} \{g(y)\}\,,
\label{eq:vel}
\end{equation} 
which, with the choice of the sin flow, means $V_f=v_0+U$.  Similarly
one can predict the asymptotic shape of the front, which is shown in
Fig.~\ref{fig:1}a. The important features are the presence of a
stationary (maximum) point in correspondence of the point where $g(y)$
has its maximum, and a cusp in its minimum. The asymptotic speed
(\ref{eq:vel}) is reached only after the transient time $t_w$
necessary to the front shape to reach its maximum length
(corrugation).  Following \cite{KBM01} we call $t_w$ as the {\it
wrinkling } time, that can be defined as the time the front width
$\mathcal{W}_f$ (i.e. the distance between the leftmost point in which
$\theta=0$ and the rightmost in which $\theta=1$) employs to pass from
the initial zero-value (indeed at the beginning the front is flat) to
the asymptotic one $\mathcal{W}_f^\ast$. For $U \gg v_0$ this time can
be estimated as
\begin{equation}
t_w\propto  L/v_0\,.
\label{eq:tw}
\end{equation}
This comes from the fact that starting from the flat profile the front
width $\mathcal{W}_f(t)$ grows in time as $2Ut$ up to the moment in
which the cusp (see Fig.~\ref{fig:1}a) is formed (see also
\cite{KBM01}). Then the growth slows down up to the stationary value
$\mathcal{W}_f^\ast$. Assuming the linear growth up to the end one may
estimate $t_w=\mathcal{W}_f^\ast/(2U)$. Further, since in the shear
flow case (where the formation of pockets of inert material is not
possible), for $U \gg v_0$, the width $\mathcal{W}_f^\ast$ is
proportional to the stationary front length $\mathcal{L}_f$, which is
linked to the asymptotic velocity by Eq.~(\ref{eq:relation}). Finally
since the latter given by (\ref{eq:vel}) one ends up with
$t_w=(L/U)(1+U/v_0)$ which reduces to (\ref{eq:tw}) for $U\gg v_0$.  In
Fig.~\ref{fig:2} we show $v_f(t)$ as a function of $t/t_w$, as one can
see with this rescaling the asymptotic speed is reached at the same
instant for systems which have different $U$ and $v_0$, as the nice
collapse of the different curves indicates (compare with the
inset). We noticed that as soon as $U/v_0 \geq 4$ $t_w \propto L/v_0$
as predicted by Eq.~(\ref{eq:tw}).  \\\indent The wrinkling time is an
inner time scale of the reaction dynamics, which is very important
when considering time-dependent flows.  In particular, here we study
the example of random shear flows (\ref{eq:shear}) with
$g(y)=\sin(2\pi y/L)$ (as in the stationary case) and random
amplitudes $U(t)$ which are chosen according to an Ornstein-Uhlembeck
process. Therefore, $U$ evolves according to the Langevin dynamics
\begin{equation}
\frac{dU}{dt}=-\frac{U}{\tau_f}+\sqrt{\frac{2 U_{rms}^2}{\tau_f}}\: \eta
\end{equation}
 where $\eta$ is a zero-mean Gaussian white
noise and $\tau_f$ defines the flow correlation time so that $\langle
U(t)U(t')\rangle= U^2_{rms} \exp(-|t-t'|/\tau_f)$.  Clearly one has to
distinguish two limiting cases: i) when the flow fluctuations are
slower than the wrinkling time: $\tau_f\gg t_w$; ii) when they are
much faster: $\tau_f\ll t_w$.\\
\begin{figure}[t!]
  \iffigs \includegraphics[scale=.55]{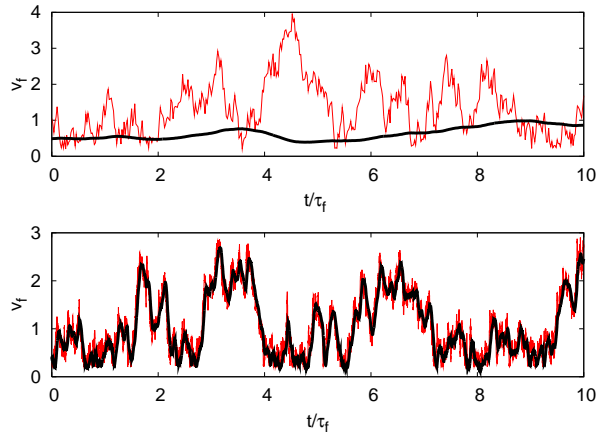}
  \else \drawing 70 90 {Fig3.eps} \fi
  \caption{\label{fig:3} Measured front velocity $v_f(t)$ (thick black
line) and the adiabatic prediction $v_0 + |U(t)|$ (solid line, red
online) versus $t/\tau_f$ for a correlated flow with $U_{rms}=1$,
$v_0=0.2$ and: (top) $\tau_f=200 (\gg t_w=31.4$), (bottom) $\tau_f (2 \ll
t_w=31.4)$.  The resolution used was $N_y=800$ in the first case and 
$N_y=3000$ in the second one.}
\end{figure}
$i)$ In this condition the front has enough time for adiabatically
adjust itself on the instantaneous flow velocity. Thus by generalizing
(\ref{eq:vel}) it is natural to expect that $v_f(t)=v_0+|U(t)|$ (as
confirmed in Fig.~\ref{fig:3}a) so that $V_f=v_0+\langle |U(t)|
\rangle$. In other words if the velocity fluctuations are slower than
the wrinkling time the front can be efficiently corrugated close to
the maximal wrinkled shape allowed by the flow and so by
(\ref{eq:relation}) can reach maximal speed.\\ $ii)$ On the other hand
in opposite limit $\tau_f\ll t_w$ the front has not time to be
maximally corrugated by the flow, and so its speed cannot reach the
maximal amplification allowed by the fluid. In this case it is not
anymore true that $v_f(t)=v_0+|U(t)|$ (see
Fig.~\ref{fig:3}b). \\\indent These effects on the propagation speed
have a counterpart in the patterns of the front. This is evident by
looking at the front shape (compare Fig.~\ref{fig:1}b and c with
a). Indeed while in the case $\tau_f \gg t_w$ at any instant the shape
of front closely resembles that obtained in the stationary case, when
$\tau_f \ll t_w$ one notices that the front length is strongly reduced
and the spatial structure complicated by the presence of more than one
cuspid.\\\indent Looking at Fig.~\ref{fig:3}b it is clear that the
reactive dynamics acts as a sort of filtering of the fluid velocity so
that not only the front speed is not enhanced at the maximal allowed
value but also its fluctuations are much decreased. In
Fig.~\ref{fig:4} we show the normalized front speed
$V_n=(V_f-v_0)/\langle|U(t)|\rangle$ and the normalized variance
$\sigma_n=\sigma_f\/(\sqrt{\langle \vert U(t) \vert^2\rangle-\langle
\vert U(t)\vert\rangle^2}$ (i.e. the standard deviation of the
$v_f(t)$ normalized by that expected on the basis of the adiabatic
process $|U(t)|$), by fixing the flow intensity $U_{rms}$ and varying
the correlation time. Note that in the limit of very long correlation
times $V_n\approx 1$ and $\sigma_n\approx 1$.  As one can see a fast
drop of the front speed and average fluctuations with respect to its
maximum value is observed when $\tau_f/t_w<1$, confirming the above
picture.\\\indent
These results along with those of Refs.~\cite{D99,A00,KBM01} confirm
the importance of time correlations in the flow in determining the
front speed. This may be relevant to more realistic flows in the light
of the above cited bending phenomenon. Indeed in turbulent flows one
has that at increasing the turbulence intensity fluctuations on
smaller and smaller scales appear. These are characterized by faster
and faster characteristic time scales. In this respect, as suggested
in by the results of this work, one may expect that in the corrugation by
these scales become less and less important, so that the average front
speed may increase less than expected.\\
\indent We conclude by noticing that it would be very interesting to test the
effects of time-correlations on the front propagation also in other
kind of laminar flows. In particular this could be performed in
experimental studies in settings similar to those of
Refs.~\cite{LMRS05} where flows of the form $\bm u(\bm r,t)=U(t)\bm
v(\bm r)$ can be easily generated with a good control of the 
time dependence of the amplitude.\\
\indent We are grateful to C. Casciola for useful discussions.  MC and AV 
acknowledge partial support by MIUR Cofin2003 ``Sistemi Complessi e
Sistemi a Molti Corpi'', and EU under the contract HPRN-CT-2002-00300.

\end{document}